\documentclass{article}
\usepackage{epsfig}
\usepackage{oljour}

\begin{document}

\journalname{zp}  

\title[en]{Theoretical study of structure and energetics 
of gold clusters with the EAM method } 

\begin{author}
\anumber{1}
\firstname{Denitsa}
\surname{Alamanova}
\institute{Physical and Theoretical Chemistry} 
\street{University of Saarland}
\number{Bld.\ B2 2}
\zip{66123}
\town{Saarbr\"ucken}
\country{Germany}
\tel{+49-681-302-3809}
\fax{+49-681-302-3857}
\email{deni@springborg.pc.uni-sb.de}
\end{author}

\begin{author}
\anumber{2}
\firstname{Valeri G.}
\surname{Grigoryan}
\institute{Physical and Theoretical Chemistry} 
\street{University of Saarland}
\number{Bld.\ B2 2}
\zip{66123}
\town{Saarbr\"ucken}
\country{Germany}
\tel{+49-681-302-4420}
\fax{+49-681-302-3857}
\email{vg.grigoryan@mx.uni-saarland.de}
\end{author}

\begin{author}
\anumber{3}
\firstname{Michael}
\surname{Springborg}
\institute{Physical and Theoretical Chemistry} 
\street{University of Saarland}
\number{Bld.\ B2 2}
\zip{66123}
\town{Saarbr\"ucken}
\country{Germany}
\tel{+49-681-302-3856}
\fax{+49-681-302-3857}
\email{m.springborg@mx.uni-saarland.de}
\end{author}

\corresponding{vg.grigoryan@mx.uni-saarland.de}

\abstract{
Using the Embedded Atom Method as developed by Voter and Chen in combination
with the {\it variable
metric/quasi-Newton} and our own {\it Aufbau/Abbau} methods, we have identified the
three most stable isomers of Au$_N$ clusters with $N$ up to 150. For
the first time clusters with tetrahedral symmetry are found to form the
ground states of Au$_{17}$ and Au$_{34}$. The Au$_{54}$ {\it icosahedron}
without a central atom and the Au$_{146}$ {\it decahedron} are found to be
particularly stable, whereas
the highly symmetric second and third Mackay icosahedra that could have been obtained 
for $N$ = 55 and
147, respectively, do not correspond to the particularly stable structures. 
The three lowest-lying isomers of Au$_{55}$ and Au$_{147}$
are low-symmetrical structures. Various structural and energetic properties are
analysed, such as stability function,  occurrence of magic-sized clusters,
construction of icosahedral and {\it fcc} shells, and cluster growth.}


\dedication{This work is dedicated to Prof.\ Dr.\ Wolf Weyrich, Physical Chemistry,
University of Konstanz, Germany, on the occasion of his 65th birthday.}

\received{x}
\accepted{x}
\volume{x}
\issue{x}
\class{x}
\Year{x}

\maketitle

\section{Introduction}

Since the first part of the 20th century it has been recognized that theoretical
studies can constitute an important ingredient of science, first through the
dedication of chairs to the field of theoretical physics and later through the
establishment of chairs in theoretical chemistry. Simultaneously, by defining
`theory' as an independent part of science, the interactions between theory and
experiment have in many cases been reduced. Persons who unite these two parts are,
therefore, of immense importance and, simultaneously, able to contribute to science
in a way that is unmatched by most colleagues. A person, occupying a chair of 
physical chemistry, and being active in both experimental and theoretical studies
of the properties of matter, is, accordingly, unique. Such a person is Wolf Weyrich,
who has made significant contributions to the understanding of the properties of a
large range of systems, going all the way from smaller molecules to extended solids.
In the later years his interests have also turned towards the properties of 
nanoparticles that somehow lie in between the materials of his earlier interests. 
Metal nanoparticles are thereby of central interest. Therefore, it is with pleasure
that we dedicate the present work to Wolf Weyrich on the occasion of his 65th birthday.

     Clusters form an important link between isolated atoms and 
molecules at one extreme and bulk solids at the other. Their large
surface-to-volume ratio gives them 
unique physical and chemical properties. On the other hand, the combination of 
finite with large size makes it difficult to characterize and analyze their properties
in detail. From a theory point of view the central difficulty lies in the determination
of the structure of the system of interest and only through comparison with experimental
information more definite statements about their properties can be made.

Gold clusters represent some of the mostly studied clusters (for a detailed
discussion, see Ref.\ \cite{pp04,bf05}). They have recently been investigated in 
connection with the synthesis of nanostructured materials and devices 
\cite{rw96,as99,hh03,lds03,hb04}. Their structural and energetic properties
have been studied with High-Resolution Electron 
Microscopy (HREM) and various spectroscopic techniques
\cite{kjt92,hh94,ap95,mma97,tgs97,cc97,kk98,bp98,leh99,vas00,mv01,nv01,hh01,
vt03,rs04,gm04,bkm04,hm04}. Literature concerning small Au$_N$ clusters 
is enriched with numerous investigations 
based on density-functional methods 
\cite{ll97,odh97,gbp99,hh99,ga00,hl00,jms00,hh02,jo02,jw02,jw03,zw03,by03,
mj04,sc04,rmo05,avw05,lx04,fr05} that are not yet capable of giving a definite
answer to the problem at what cluster size the structural 2D -- 3D transition
occurs. Recent studies combining theory and experiment
\cite{hy03,sg02,ff02,mn05} show that the gold clusters are planar at least up
to $N$ = 7 for Ref.\ \cite{sg02}, or up to $N$ = 12, according to
H\"akkinen \cite{hy03} and Furche {\it et al.} \cite{ff02}. 

However, global
structure optimization is difficult when using {\it ab initio} methods already at very
small cluster sizes. Nevertheless, some studies in this direction exist.
Thus, the authors of Ref.\ \cite{odh97,mj04} performed
density functional calculations on clusters containing more than 30 atoms,
relaxing selected high-symmetric configurations. Alternatively, the global
optimizations of larger clusters are all based on approximate methods like
molecular dynamics 
\cite{fe91,ha98,wdl98,clc98,clc99,yc01,sp01,scl02,fb02,jr04,ul04,yw04} and
semiempirical potentials like the EAM 
\cite{cc97,clc97,rb99}, Sutton-Chen \cite{dw98}, Murrell-Mottram
\cite{ntw00,ntw02}, or the many-body 
Gupta potential \cite{ilg96,ilg98,ilg99,km99,tl00,sd02}. Using these methods,
unbiased structure optimizations  
were performed up to the 80-atom cluster. Medium-sized
clusters ( $ 80 \le N \le 150$) have hardly been 
studied. Besides the first-principles study of H\"aberlen {\it
et al.} \cite{odh97} and the EAM calculations by Cleveland {\it et
al.} \cite{cc97,clc97,rb99}
considering particular structural motifs, there exists essentially no further
investigation on the clusters in this size range. 

In most of the studies, special attention is paid to the so-called
`magic-numbered' clusters, that 
possess closed electronic and/or geometric shells. Various studies on the
smallest `magic' cluster Au$_{13}$ have identified the formation of an
icosahedron \cite{ll97,odh97,dw98,ntw00,km99,sd02}. Only the authors of
Ref.\ \cite{jo02,jw02} found a 
disordered structure as the lowest-lying isomer for this cluster size. On the other
hand, semiempirical potentials \cite{clc97,dw98,ntw00,sd02} and the
density functional study by H\"akkinen {\it et al.} \cite{hh99} on the
Au$_{38}$ cluster predict the truncated octahedron to be the 
global minimum for this cluster size. However, on the basis of first-principles
and Gupta potential calculations, the 
authors of Ref.\ \cite{ilg99,tl00} state that a disordered structure is
actually lower in energy than the symmetric. Ultimately, it may be suggested that 
the obtained structure depends sensitively on the type of the
potential, since by using another 
form of the same potential, Darby {\it et al.} \cite{sd02} found a truncated
octahedral structure to be the global minimum 
of Au$_{38}$. The situation is more clear for the Au$_{55}$ and Au$_{75}$
clusters, where a disordered 
structure \cite{dw98,ilg96,ilg98,km99,tl00,sd02} or a Marks decahedron
(m-D$_{5h}$) \cite{ilg98,km99} seem to constitute the global minima. 

Several groups have performed calculations on larger clusters by minimizing the
total energy of initially chosen
symmetric structures, although it may be feared that the structures may not be those of
the global total-energy minima. Moreover, only few cluster sizes 
were studied --- the octahedral Au$_{79}$ \cite{clc97,dw98} and Au$_{140}$
\cite{clc97}, the decahedral 
Au$_{101}$, Au$_{116}$, and Au$_{146}$ \cite {clc97}, and the icosahedral
Au$_{147}$ \cite 
{odh97,clc97}. The structures and energetics of the clusters between these high
symmetrical ones remain scarcely investigated.

The purpose of the present study is to carry out unbiased calculations on small and
intermediate gold clusters, 
and, subsequently, to investigate the occurrence of magic clusters and the growth 
patterns, especially for larger clusters. To our knowledge, two earlier studies 
on gold clusters with the same version of EAM have been performed previously, i.e., the
work of Rey and 
coworkers \cite{cr93,cr94} and that of Sebetci and G\"uven\c{c} \cite{as04}. 
The first study considered only the energetics and stability of small gold
clusters ($2 \le N \le 23$). Sebetci {\it et al.} used a 
basin-hopping Monte Carlo minimization approach to find the global minima of
Al$_N$, Au$_N$, and 
Pt$_N$ clusters with $N \le 80$. The total energies, point groups, and
structural assignments were presented.

In the present study the structure and energetics of the three most stable
isomers of small and 
medium-sized Au$_N$ clusters with $2 \le N \le 150$ have been determined for
each cluster size by 
using a combination of the embedded-atom method in the version 
of Voter and Chen \cite{vo87,vo93,vo95}, the {\it variable metric/quasi-Newton}
method, 
and our own {\it Aufbau/Abbau} method. The paper is organized as follows. 
In Sec.\ 2 we briefly outline the embedded-atom method, and in Sec.\ 3 we
present our structural-determination methods. The main results are given in
Sec.\ 4, and a brief summary is offered in Sec.\ 5.

\section{The Embedded Atom Method}

The main idea of the EAM was initially presented by Daw, Baskes, and Foiles (DBF)
\cite{da83,da84,fo86} in 1983--1986, and since then the generality of 
the functions of the EAM 
of DBF has been successfully tested through numerous applications to different
systems of metals and alloys, including defects, surface and 
interface structures, surface and bulk phonons, etc. In a previous paper
\cite{our} we reported results for the global minima of Ni$_N$, Cu$_N$, and
Au$_N$ clusters with up
to 60 atoms, obtained with two different versions of the EAM. There we discussed the 
incapability of the so-called DBF version of 
EAM to describe properly the properties of the smallest gold clusters,
which could be related to the parameterization of the potential 
only to bulk properties. The version developed by Voter and Chen
\cite{vo87,vo93,vo95} takes into account also the properties of the dimer,
which makes this method more suitable for the description of the
smallest clusters.
Accordingly, in this study we use this version of the EAM for the calculation of
the total energy of a given cluster. 

The principle of the method is to split the total energy of the
system into a sum over atomic energies:
\begin{equation}
E_{tot}=\sum_{i}^{N}E_i.
\label{1}
\end{equation}
The embedding energy is obtained by considering each atom as an impurity
embedded into a host 
provided by the rest of the atoms. The electron-electron interaction is
presented as sums of short-ranged, pair potentials. Accordingly,
\begin{equation}
E_{tot}=\sum_i F_i(\rho_i^h)+{\frac{1}{2}\sum_{i\ne j}\phi_{ij}(r_{ij})}
\label{2}
\end{equation}
where $\rho_i^h$ is the local electron density at site $i$, $F_i$ is the
embedding energy, i.e., 
the energy required to embed an atom into this density, and $\phi_{ij}$ is a
short-range potential between atoms $i$ and $j$ separated by distance $r_{ij}$.
The pair potential, according to the Voter-Chen version, is taken to be a Morse
potential,
\begin{equation}
\phi(r)=D_M[1-e^{-\alpha_M(r-R_M)}]^2-D_M
\label{3}
\end{equation}
where the three parameters, $D_M$, $R_M$, and $\alpha_M$, define depth, position
of the minimum, 
and a measure of the curvature at the minimum, respectively. The local density
at site $i$ is assumed being a superposition of atomic electron densities:
\begin{equation}
\rho_i^h=\sum_{j \,(\ne i)}\rho_j^a(r_{ij})
\label{4}
\end{equation}
where $\rho_j^a(r_{ij})$ is the spherically averaged atomic electron density
provided by atom $j$ 
at the distance $r_{ij}$. The density function is taken as the density of a
hydrogenic $4s$ orbital:
\begin{equation}
\rho(r)=r^6[e^{-\beta r}+2^{9}e^{-2\beta r}]
\label{5}
\end{equation}
where $\beta$ is an adjustable parameter. Because $r^6e^{-\beta r}$ turns over
at short $r$, the 
second term has been added to maintain the monotonically decreasing character
of $\rho(r)$ at 
shorter $r$. This $4s$ orbital density, appropriate for Ni and Cu, also works
well for gold.
To ensure that the interatomic potential and its fist derivatives are
continuous, both $\phi(r)$
and $\rho(r)$ are cut off at $r$=$r_{\rm cut}$. 
In the fitting procedure, the five parameters defining $\phi(r)$ and $\rho(r)$
($D_M$, $R_M$, 
$\alpha_M$, $\beta_M$ and $r_{\rm cut}$) are optimized  by minimizing the
root-mean-square deviation 
between the calculated and reference properties of different selected systems like
molecules, surfaces, solids, and defects. Because
$F(\rho^h)$ is 
redefined for each choice of the parameters, the potential always gives perfect
agreement with 
experimental values such as $a_0$, $E_{\rm coh}$ and the bulk modulus $B$. The
reference properties are
the three cubic elastic constants ($C_{11}$, $C_{12}$ and $C_{44}$), the
unrelaxed vacancy 
formation energy ($E_{vac}^f$), and the bond length ($R_e$) and bond strength
($D_e$) of the 
diatomic molecule. The values of $\rho_i^a$, $\phi_{ij}$ and $F_i(\rho_i)$ that
were used by the Voter-Chen version, are 
available in numerical form for Ni, Pd, Pt, Cu, Ag, Au and Al. 

Our reason for choosing the EAM was dictated by the good agreement to 
experiment, as well as to first principles 
calculations, and last but not least by the high computational efficiency
allowing one to investigate clusters with more than 100 atoms without severe
constraints on the initial geometry, which is 
impossible with first principles methods.
In a previous work \cite{our} we performed
calculations on smaller gold clusters with $2 \le N \le 60$ atoms comparing the
EAM of Daw, Baskes, and Foiles (DBF) and the version used in this work, and we
found that the DBF overestimated the
binding energy of the dimer by 209{\%} and underestimated the bond distance by
37{\%}. For comparison, the EAM of Voter and Chen gives dimer binding energy
corresponding to 99.6{\%} of the experimental value, and bond distance that is
92.2{\%} of the experimental value. For this reason we 
chose to work with the Voter-Chen version that describes correctly the dimer
properties. 

\section{Structure optimization}

Using expression (\ref{1}) we can calculate the total energy of
any cluster with any structure as a function of structure, i.e., of the atomic coordinates
$\{ {\vec R}_i\}$, $E_{\rm tot}({\vec R}_1,{\vec R}_2,\, \dots \, ,{\vec
R}_N)$. 
In order to obtain the closest local total-energy minimum we use the 
{\it variable metric/quasi-Newton} method \cite{numer92}.

For searching the global minima we have developed our own {\it Aufbau/Abbau}
method that is described in details in previous works \cite{cpl03,prb04}. It
consists of the following steps:

1) We consider two cluster sizes with $N$ and $N+K$ atoms with $K\simeq 5-10$. 
For each of those we randomly generate and relax a large set of structures,
from which those with the lowest total energy are selected. 

2) One by one, each of the $N$ atoms is displaced randomly, and the closest
local minima is determined. If the new structure has a lower total energy than
the original one, this new one is kept, and the old one discarded. This is
repeated approximately $1000-2000$ times depending
on cluster size. 

3) This leaves us with two `source' clusters, Au$_N$ and 
Au$_{N+K}$ with their lowest total energies. One by one an atom is added 
at a random position to the structure with $N$ atoms 
(many hundred times for each size), and the structures are relaxed. In
parallel, 
one by one an atom is removed from the structure with $N+K$ atoms --- for each 
intermediate cluster with $N'$ atoms we consider all $N'+1$ possible 
configurations, that one can obtain by removing one atom from the Au$_{N'+1}$
cluster. 
From the two series of structures for $N\le M\le N+K$ those structures of the 
lowest energies are chosen and these are used as seeds for a new set of 
calculations. First, when no lower total energies are found in the two
sets of calculations, it is assumed 
that the structures of the global-total-energy minima 
have been identified. 

\section{Results and Discussion}

\subsection{Small gold clusters}

Like all other semiempirical potentials, the one of the embedded-atom method
does not include explicitly the electrons and their orbitals. 
Therefore, such a potential tends to prefer high-symmetry, compact structures,
whereas structures of lower symmetry that can be explained through electronic
effects are not found. As a result, for the smallest gold clusters, 
where spin-orbit interactions
play an important role, our global-minima structures are compact (see Table
\ref{tab01}), and the planar structures (that are believed to be those of the
true total-energy minima) only metastable. In the size range $N$
= 4 -- 7 first principles studies obtain these 3D configurations as
higher-lying isomers, which can serve as an example of how the inclusion of
electronic effects can change the energetic ordering of the isomers. On the
other hand, the addition of electronic effects in the semiempirical potentials would
restrict their use only to small and relatively larger clusters with pre-chosen structures. An
appropriate choice in this respect could be the Density Functional Tight-Binding methods (DFTB)
that include explicitly the electrons and are computationally more efficient 
than the common density functionals. Actually, in a recent study \cite{cole} we demonstrated
the important role that the electronic effects can play for the
binding energy and the stability functions. However, except for $N$ = 4, where
a rhombus was the lowest-lying isomer according to the DFTB method, all the
ground state structures for the smallest gold clusters Au$_5$ -- Au$_9$ had 3D
shapes. On the other hand, even the most recent density-functional studies are
still not in agreement at which cluster size the structural transition 2D -- 3D
occurs. According to the LDA study of Wang {\it et al.} \cite{jw02}, it is the
pentagonal bipyramid that forms the global minimum of Au$_7$. Remacle and
Kryachko \cite{fr05} suggested that gold clusters are planar at least up to
$N$ = 9, while Walker \cite{avw05} predicted that the transition occurs at
Au$_{11}$. Using ion mobility measurements and {\it ab-initio} molecular
dynamics Kappes {\it et al.} \cite{sg02,ff02} found that the 3D transitions
occur at Au$_{12}^-$ and Au$_8^+$. The same group studied the adsorption of CO
on isolated gold cluster cations in the size range $N$ = 1 -- 65. The smallest
clusters with $4 \le N \le 6$ as well as Au$_8$ were found to be planar, while for
Au$_7$ the global minimum was a 3D structure, but not a bipyramid, in contrast
to the results of Wang {\it et al.} \cite{jw02}. In a combined experimental and
theoretical study H\"akkinen and coworkers \cite{hy03} confirmed the 2D -- 3D
transition at Au$_{12}^-$, however, Xiao and Wang \cite{lx04} suggested that for the
neutral clusters this transition occurs first at Au$_{15}$. In most of
the cases the planar structures are competing with 3D isomers, and the
energetic differences are insignificant, which in turn means that the
ordering of the isomers depends strongly on the used functional and the
starting conditions. For example, H\"akkinen {\it et al.} \cite{hh02} compared
the global minima of relativistic and nonrelativistic Au$_7^-$ clusters and
found that for the nonrelativistic gold the lowest-lying isomer was a capped
octahedron that corresponds to our second isomer for this cluster size. At
larger cluster sizes the potential
used in this study yields results in agreement with density-functional and
experimental studies. The study of H\"akkinen {\it et al.} \cite{hh99} on the
Au$_{38}$ cluster predicted the truncated octahedron to be the global minimum,
and a recent experiment \cite{hm04} showed that the Au$_{55}^-$ cluster most
probably is not an icosahedron, but a structure with a low symmetry, in
agreement with our results.

In summary, we can conclude that although our results for the smallest gold
clusters correspond to higher-lying isomers within the first principles
methods, due to the lack of electronic effects, our method is sufficiently
accurate in describing the larger gold clusters with $N>9$, where most probably
the planar structures begin to compete with 3D configurations.

\subsection{Energetic properties}

The high stability of the so-called `magic-numbered' clusters has become a
subject of great interest 
in connection with its relevance in the medicinal and the colloidal
chemistry, as well as in the production of catalysts and high-tech nanomaterials.

\begin{figure}                                 
\begin{picture}(14,08)                         
\put(0,0){\psfig{file=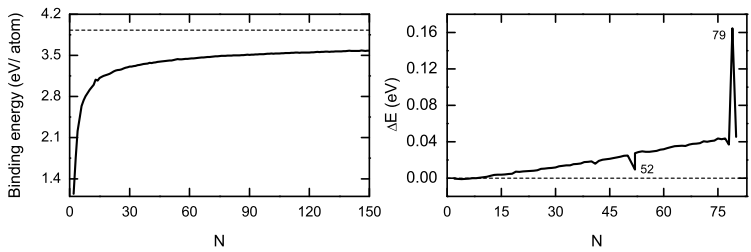,width=11cm}}
\end{picture}                                  
\caption{On the left we show the binding energy per atom as a function of size for
the energetically                              
lowest isomers of $N$ up to 150 with the dashed line giving the {\it bulk} value; on
the right is displayed the difference between the         
total energies of the lowest-lying isomers with up to 80 atoms obtained by
Sebetci {\it et al.} and those found here, using different structure-optimization
methods but the same embedded-atom approach.}   
\label{bind}                                   
\end{figure}

In Fig.\ \ref{bind} we show the binding energy per atom for the
global-minima structures, as well as the difference between the 
total energies of the lowest-lying isomers
obtained by us and those found by Sebetci {\it et al.} \cite{as04} using exactly
the same potential for the interatomic interactions. One 
can see that the latter difference increases almost linearly with the number of atoms
and has its maximum at $N$ = 
79, where we obtained a truncated octahedron in contrast to the structure with
D$_{3h}$ symmetry found by Sebetci and G\"uven\c{c}. Except for few cases, 
the total-energy difference is marginal (about 5 meV/atom, which may be due to
numerical differences), giving support for the quality of both theoretical 
approaches in optimizing the structure. 
At $N$ = 52 they obtained an uncentered
icosahedron-like structure with C$_{2h}$ symmetry, that lies energetically
between our second and third lowest isomers. 

\begin{figure}                                 
\begin{picture}(10,10)                         
\put(0,0){\psfig{file=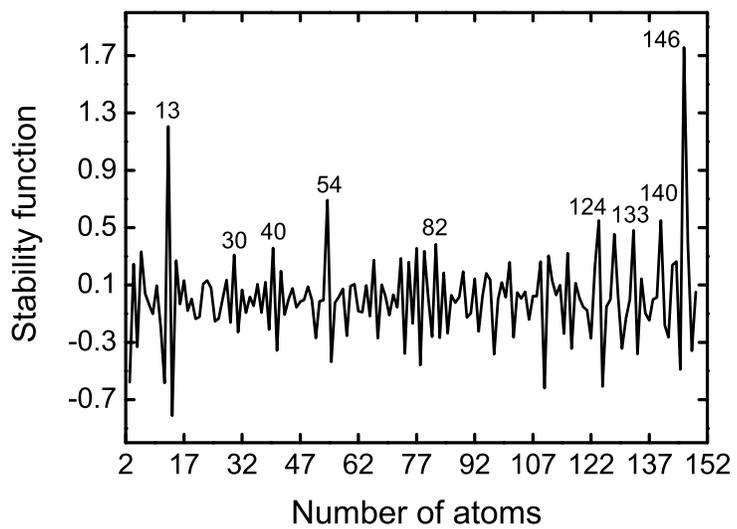,width=11cm}}
\end{picture}                                  
\caption{The stability function as a function of cluster size.}
\label{stability}                              
\end{figure}

In order to identify the particularly stable clusters we have considered the following 
criteria. The clusters can be considered as very stable if their binding energy
per atom is much 
larger than that of the two neighboring clusters. This can be quantified
through the stability function,
$E_{\rm tot}(N+1.1)+E_{\rm tot}(N-1.1)-2E_{\rm tot}(N.1)$, where $E_{\rm
tot}(N.k)$ is the
total energy of the energetically $k$-lowest isomer of the Au$_N$ cluster. This
function, that has maxima for particularly stable clusters, is shown in Fig.\
\ref{stability}.
Here we can identify a large number of particularly stable clusters, i.e.,
so-called magic
clusters. These are found for 
$N$ = 4, 6, 10, 13, 15, 17, 23, 28, 30, 36, 38, 40, 42, 45, 49, 54, 58, 61, 
64, 66, 68, 73, 75, 77, 79, 82, 84, 89, 92, 95, 101, 109, 111, 116, 118, 124,
128, 133, 135, 140, 144, and 146. The most pronounced peaks (marked in the figure) 
occur at $N$ = 13, 30, 40, 54, 75, 79, 82, 124, 133, 140, and 146. In agreement with
Sebetci and G\"uven\c{c} \cite{as04}, 
the 54-atom icosahedron without a central atom is found to represent a 
magic-numbered cluster, whereas the Au$_{55}$ cluster does not. 
The latter possesses a distorted icosahedral structure with C$_{3v}$
symmetry, lying 0.374 eV lower than the perfect icosahedron, 
2.9 eV lower than the decahedron, and 3.27 eV lower than 
the cuboctahedron. In our study, all the three lowest-lying
isomers of Au$_{55}$ have lower energy than 
the symmetric structures, in agreement with previous studies where disordered
configurations were 
found as global minima for Au$_{55}$ \cite{ilg96,ilg98,tl00,sd02}. For Au$_{38}$
and Au$_{75}$, a 
cuboctahedron \cite{hh99,dw98,ntw00,sd02,as04} and a Marks decahedron
\cite{dw98,ilg98,km99,as04} were 
obtained, in agreement with first-principles and semiempirical studies.
However, two studies employing 
the many-body Gupta potential identified amorphous structures as those of the global minima
of Au$_{38}$ 
\cite{ilg99,tl00}, which is most probably due to the parameterization of the
potential, since Darby {\it et al.} \cite{sd02} found an octahedron as the
lowest-lying isomer by using another version of the same potential. 

Another striking result of our study is that the 146-atom Marks decahedron represents
a peak in the stability function, whereas 
the Au$_{147}$ icosahedron does not. According to our study, the third Mackay
icosahedron lies 2.89 eV lower than 
the cuboctahedron, which in turn is 2.53 eV lower than the decahedron, but 0.37 eV
higher than a disordered structure 
with partly decahedral construction. To our knowledge, this is the first study
predicting a disordered global minimum for the Au$_{147}$ cluster.

\begin{figure}                                 
\begin{picture}(10,10)                         
\put(0,0){\psfig{file=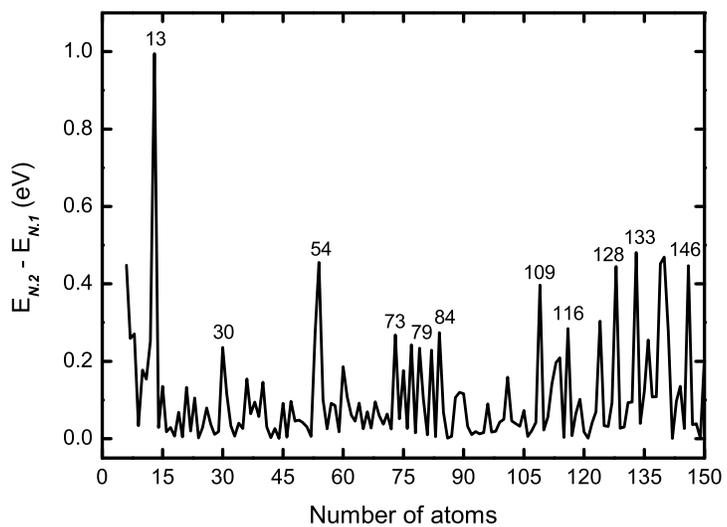,width=11cm}}  
\end{picture}                                  
\caption{The total-energy difference between the two energetically lowest
neighbouring isomers as a function of cluster size.}
\label{diff}                                   
\end{figure}

According to our other criterion for a particularly stable cluster, such a
cluster occurs if the energy difference between the two energetically lowest
isomers $E_{\rm tot}(N.2)-
E_{\rm tot}(N.1)$ is large. This energy difference is shown in Fig.\
\ref{diff}, and comparing to Fig.\ \ref{stability} we can see that many of the
clusters that are particularly stable according to the first criterion are
stable also according to the second one. 

\subsection{Structural properties}

In this subsection, instead of discussing in particular the structures of the
individual clusters, we shall introduce different quantities that are devised
to reduce the available information to some few key numbers. The theoretical
background of the descriptors used in this subsection was introduced by us in a
previous work \cite{prb04}.

\begin{figure}                                 
\begin{picture}(10,12)                         
\put(0,0){\psfig{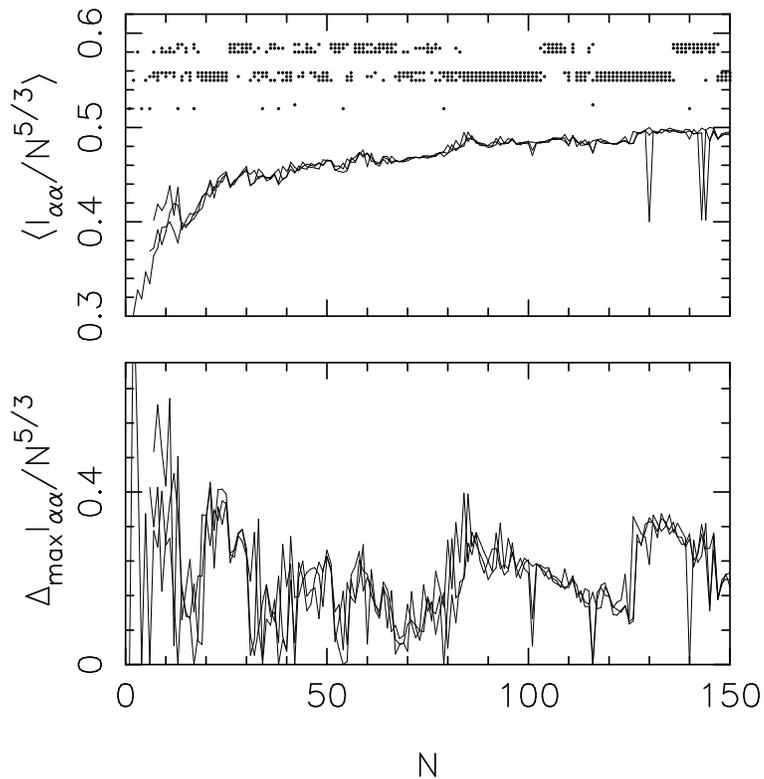}}        
\end{picture}                                  
\caption{Different properties related to the eigenvalues $I_{\alpha\alpha}$ of the
matrix with the moments of inertia. In the upper panel                                
we show the average value together with points indicating whether clusters with
overall spherical shape (lowest set of rows), overall cigar shape (middle set of rows),
or overall lens shape (upper set of                            
rows) are found for a certain size. Moreover, in each set of rows, the lowest
row corresponds                                
to the energetically lowest isomer, the second one to the energetically
second-lowest                                  
isomer, etc. In the lower panel we show the maximum difference of the
eigenvalues for the three different isomers.}  
\label{shape}                                  
\end{figure}

The shape analysis, based on the eigenvalues of the matrix with the moments of
inertia and whose results are shown in Fig.\ \ref{shape}, separates the clusters into being
overall spherical, more cigar-like shaped, or more lens-like shaped.
One can see that only few clusters have a spherical shape (these are
found for the energetically lowest isomer for $N$ = 4, 6, 13, 17, 34, 38, 54,
79, and 140, and
for the next one for $N$ = 42 and 116), all of them corresponding to
high-symmetrical isomers
(cf.\ Table \ref{tab01}) and, for the lowest-energy isomer, most of them to the
class of magic clusters. 
It is interesting that the average value follows more or less 
the same curve for all the three isomers, with some deviations at
$N$ = 130, 146, and 147. Also the
largest differences show a similar behaviour, except for some few cases mainly
for $N$ below 40 and between 80 and 85. Therefore, except when the eigenvalues
are all very similar (which occurs
for $N$ around 50, 70, 100, 116, and 140), the overall shape
(i.e., lens- or cigar-like) is the same for all three isomers.

\begin{figure}
\begin{picture}(10,8)
\put(0,0){\psfig{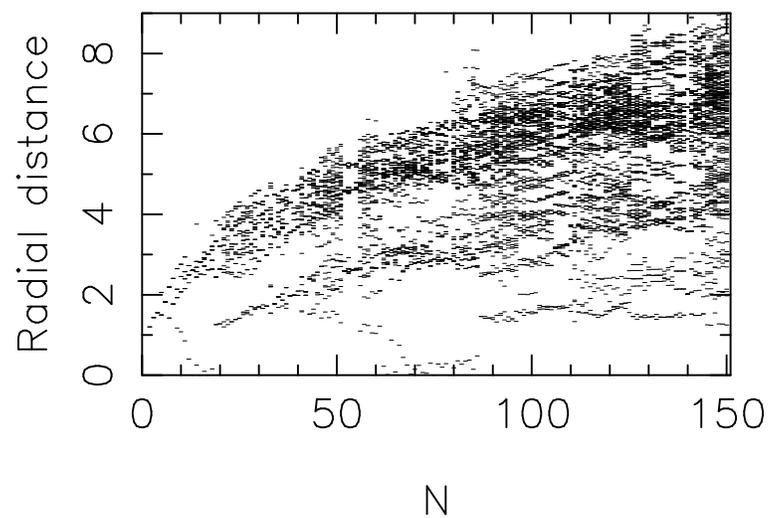}}
\end{picture}
\caption{The distribution of radial distances (in \AA) for the lowest-lying
isomer as a function of cluster size. Each
small line represents (at least) one atom with that radial distance.} 
\label{radial}
\end{figure}

The construction of atomic shells can be easily seen from the distribution of
radial distances (i.e., the distance for each individual atom to the center of mass) shown
in Fig.\ \ref{radial} for the ground state structures as function of the
cluster size. Up to $N$ around 50, no trends can be identified, with an exception 
around $N=13$. But for 
$N$ just above 50 a clear tendency towards shell construction can be seen for
the first isomer. This corresponds to the formation of the Au$_{54}$
icosahedral cluster. Also for $N$ close to 110 and around 140 shell
constructions for the lowest-lying isomer are observed. In the latter case,
this corresponds to the formation of an octahedron. The radial distributions
for the second and the third isomers are not shown, as they are quite similar
to that for the first isomer. Particular shell constructions are found
only for highly symmetrical clusters corresponding to $N$ = 42, 48,
80, 101, 116, and around Au$_{130}$ for the second isomer, and around $N$ = 40,
60, 116, and 130 for the third isomer.

\begin{figure}
\begin{picture}(10,15)
\put(0,0){\psfig{file=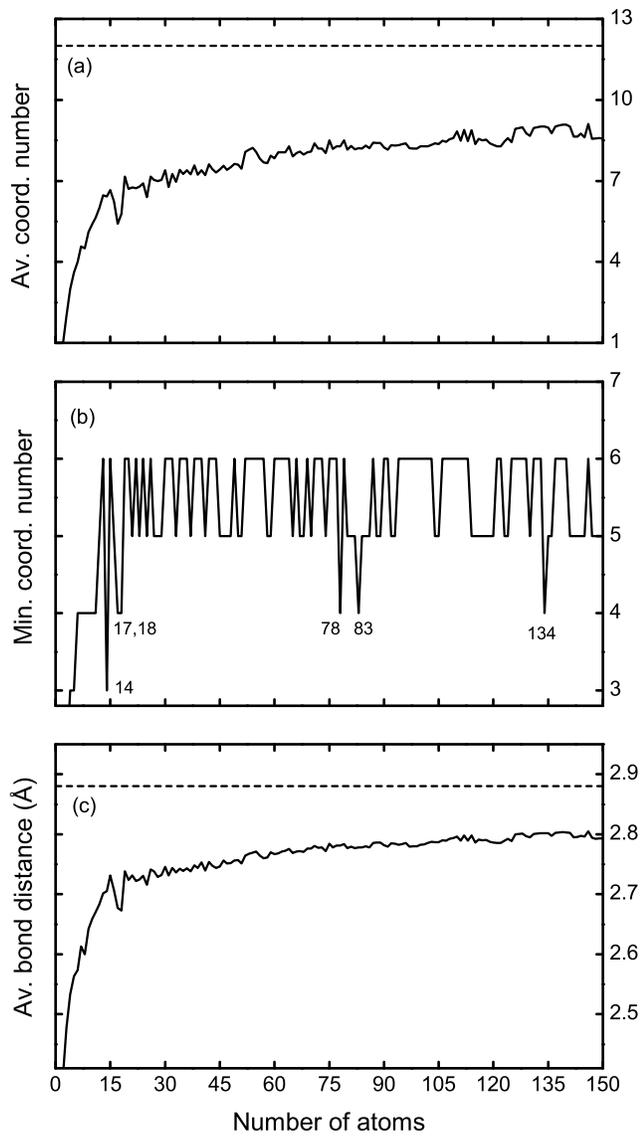,width=10cm}}
\end{picture}
\caption{(a) the average coordination number, (b) 
the minimum coordination number, and (c) the average bond length as 
functions of cluster size. The dashed lines in (a) and (c) show the 
corresponding bulk values for gold.}
\label{koord}
\end{figure}

In Fig.\ \ref{koord} we show the average and minimal coordination numbers
and the average bond lengths of the clusters. We define two atoms as being
bonded if their interatomic distance is less than 3.49 \AA, 
which is the average value between the nearest-neighbour distance  (2.89\
{\AA})
and the next-nearest-neighbour distance (4.08\ {\AA}) in bulk Au. Moreover, we
distinguish between inner atoms with a coordination number of 12 or larger and
surface atoms with a coordination number less than 12. 

Fig.\ \ref{koord}(a) presents the average coordination number as a function of
$N$. A saturation towards
the bulk limit of 12 is seen, although one has to remember that even for the
largest cluster
of our study 94 out of 150 atoms are characterized as surface atoms. Also, the
function increases in general with the size of the system, with
oscillations in particular for the clusters with $N$ = 17 and 18, which is
due to the formation of a tetrahedron for Au$_{17}$, and a structure with
C$_{4v}$ symmetry at $N$ = 18, respectively. The latter has already earlier been obtained
with the EAM method (see Ref.\ \cite{as04}), but it is the first time
that a tetrahedral configuration is found for the Au$_{17}$ cluster. 

The minimum atomic coordination for each cluster size is shown in Fig.\
\ref{koord}(b). The existence of low-coordinated atoms, i.e with coordination
numbers of 3 or 4, could point to the occurrence of a cluster growth, where
extra atoms are added to the surface of
the cluster, whereas higher coordination numbers could indicate a growth
where atoms are inserted inside the 
cluster, or, alternatively, upon a strong rearrangement of the surface atoms.
The latter is the case for the gold clusters, with few exceptions at $N$ = 14, 17,
18, 78, 83, and 134, where lower coordinations are found. The lowest
coordination corresponding to Au$_{14}$ is in connection with the
formation of an icosahedron plus one additional atom on the surface. At $N$ =
17 and 18, some structural changes take place, as discussed above. Au$_{78}$ and
Au$_{83}$ correspond to structures with a decahedral motif capped with one
additional atom. This is also the case for Au$_{134}$ where the C$_{2v}$
symmetry of the decahedral structure corresponding to $N$ = 133 is lowered by the
addition of an atom to the surface. 

Fig.\ \ref{koord}(c) shows the average bond length as a function of the cluster
size. The dashed line corresponds to the bulk value of 2.89\ {\AA}. The
average bond length for all the structures is smaller than the bulk value,
especially for Au$_{17}$ and Au$_{18}$, where more compact structures are
formed. However, this property approaches the bulk value faster than the
average coordination number.

One important issue in many of the molecular dynamics studies on gold clusters is
to identify how the clusters grow and if the cluster with $N$ atoms could be
derived 
from the one with $N-1$ atoms simply by adding one atom. In order to quantify
this possibility we use the concept of similarity functions,
introduced by us earlier \cite{cpl03,prb04}.

\begin{figure}
\begin{picture}(10,10)
\put(0,0){\psfig{file=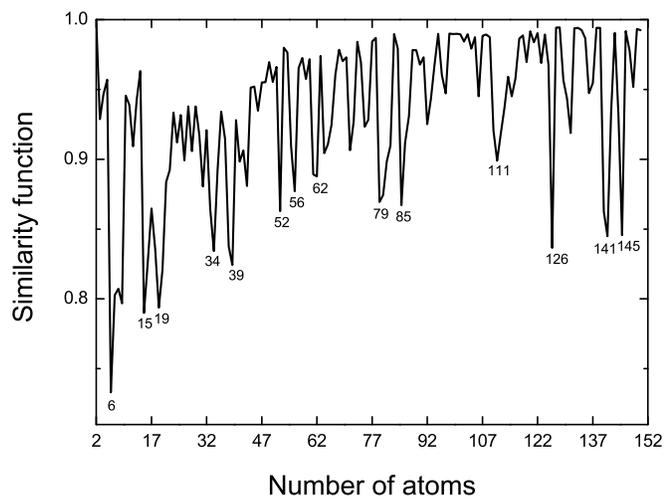,width=10cm}}
\end{picture}
\caption{The similarity function as function of cluster size. It describes
whether the cluster with $N$ atoms is similar to that of $N-1$ atoms plus an
extra atom.}
\label{similar}
\end{figure}

The similarity function $S$, shown in Fig.\ \ref{similar}(a), approaches 1 if the
Au$_N$ cluster is very similar to the Au$_{N-1}$ cluster plus
an extra atom. We see indeed that for $N$ up to around 50,
$S$ is significantly different from 1, confirming that in this range
the growth is complicated. The most pronounced peaks occur at $6<N<9$,
$15<N<20$, 34, 38, 39, 52, 56, 79, 80, 85, 111, 126, 140, 141, and $145< N
<147$. Many of these correspond to highly symmetrical clusters, however some of
the clusters with larger peaks ($N$ = 39, 56, 62, 85, 111, 126, 141, and 145)
have lower symmetry. The octahedral Au$_{38}$ and the low-symmetrical
Au$_{39}$ are structurally very different from their $N$-1-atom neighbours.
Au$_{56}$ marks the end of the icosahedral shell built between Au$_{52}$ and
Au$_{55}$, and the clusters resume their disordered growth. The octahedral
Au$_{61}$ is followed by the disordered Au$_{62}$, and the decahedral
Au$_{85}$ comes after the disordered Au$_{84}$. Between the decahedral
Au$_{110}$ and Au$_{112}$ lies the disordered Au$_{111}$. The addition of one
atom to the disordered Au$_{125}$ leads to the formation of an unfinished but
regular decahedron at $N$ = 126. The decahedral Au$_{141}$ comes immediately
after the octahedron corresponding to $N$ = 140. Although Au$_{144}$ has
partly decahedral construction, its $N$+1-atom neighbour is disordered. It
seems that for each cluster size there is a rearrangement of the gold atoms,
and no particular growth motif can be identified. This, in turn, means that the
cluster growth is very complicated and it is difficult to consider it as an
one-by-one atom addition.

\begin{figure}
\begin{picture}(12,10)
\put(0,0){\psfig{file=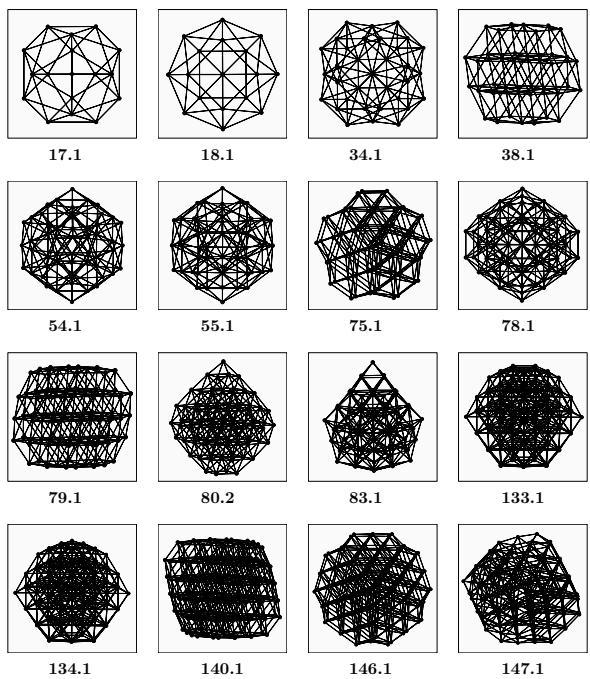,width=12cm}}
\end{picture}
\caption{Some Au$_N$ clusters with high or peculiar symmetry.}
\label{pict}
\end{figure}

Finally, some selected, high-symmetry clusters are shown in Fig.\ \ref{pict}.

\section{Summary and Conclusions}

We have determined the three energetically lowest isomers of gold 
clusters in the range $2 \le N \le 150$ by using a combination 
of the embedded-atom method in the version of Voter and Chen
(for the calculation of the total energy for a given structure), the 
{\it variable metric/quasi-Newton} method (for the determination of the closest
total-energy minimum), and our own {\it Aufbau/Abbau}
method (for the determination of the global total-energy minimum). Although the
calculations
provide a large amount of information for each individual cluster, instead of
discussing each cluster separately, we focused on identifying general trends
such as total energy per
atom, overall symmetry and shape, average bond length and coordination number,
and similarity with $N-1$-atom clusters.

The version of EAM used in the present
calculations is parameterized to bulk, as well as to the dimer properties,
which allows it to describe properly the properties of the smaller gold
clusters. 

This study predicts a number of particularly stable clusters, i.e.,
`magic-numbered' clusters that in
many cases are in agreement with results obtained by first principles and other
semiempirical studies when such exist, but the advantage of our study is that
the structures were obtained by using a completely
unbiased approach. These magic numbers were clearly visible both in the
`stability function'
and in the total-energy difference between the energetically lowest and
higher-lying isomers.

We also found that even for our largest clusters the binding energy per atom has
still not converged to the bulk limit. Similarly, the average coordination
number is far from the bulk value, but higher than for nickel clusters, where
several structures with shell constructions and corresponding low coordination
numbers were formed \cite{prb04}. The average bond distance for gold has not
reached the bulk value, due to the rearrangement of the atoms for each
cluster size that leads to the formation of very compact structures.

The shape analysis showed that roughly spherical clusters corresponded mainly
to the energetically lowest isomer, but in some cases also to the
second-lowest one, and that these often belong to particularly stable
structures. 

By analysing the distribution of radial distances as a function of the cluster
size we could
identify a region with $N$ around 55, where a shell construction was formed.
Comparing to previous results for nickel clusters \cite{prb04}, where clear
shell
constructions were formed at $N$ around 13, 55, and 147, here the atoms
rearrange for each global minimum, and therefore particular shell constructions
can not be observed. The similarity function also points to the lack of regular
growth.

\section{Acknowledgments}
We gratefully acknowledge {\it Fonds der Chemischen Industrie} for the very 
generous support.
 This work was supported by the SFB 277 of the University 
of Saarland and by the German Research Council (DFG) through project
Sp439/14-1.
Finally, it is a pleasure to dedicate this work to Wolf Weyrich who, through
his work, demonstrated to one of the present authors (MS) the importance of 
combining theory and experiment.

\begin{table}
\begin{center}
\caption{Point groups of the optimized gold clusters. $N.k$ marks the energetically $k$-lowest isomer
of the Au$_N$ cluster.}
\begin{tabular}{cccc|cccc|cccc}
\hline\hline
  N & N.1 & N.2 & N.3 & N & N.1 & N.2 & N.3 & N & N.1 & N.2 & N.3 \\
\hline
  &  & &                                        &51 & C$_{\rm 1}$ & C$_{\rm 1}$ & C$_{\rm 1}$    & 101 & C$_{\rm 2}$ & D$_{\rm 5h}$ & C$_{\rm 1}$ \\  
2 & D$_{\rm \infty h}$ & &                      &52 & D$_{\rm 5d}$ & C$_{\rm 2v}$ & C$_{\rm 2v}$ & 102 & C$_{\rm 1}$ & C$_{\rm 1}$ & C$_{\rm 1}$  \\      
3 & D$_{\rm 3h}$ & &                            &53 & C$_{\rm 5v}$ & C$_{\rm 3v}$ & C$_{\rm 1}$  &  103 & C$_{\rm 2}$ & C$_{\rm 2}$ & C$_{\rm 1}$ \\        
4 & T$_{\rm d}$ & &                             &54 & I$_{\rm h}$ & C$_{\rm 1}$ &  C$_{\rm s}$   & 104 & C$_{\rm 1}$ & C$_{\rm 2}$ & C$_{\rm 1}$\\            
5 & D$_{\rm 3h}$ & &                            &55 & C$_{\rm 3v}$ & C$_{\rm s}$ & C$_{\rm 1}$   & 105 & C$_{\rm 2}$ & C$_{\rm s}$ & C$_{\rm 2}$ \\                     
6 & O$_{\rm h}$ & C$_{\rm 2v}$ &                &56 & C$_{\rm s}$ & C$_{\rm 2v}$ & C$_{\rm 2}$   & 106 & C$_{\rm 3v}$ & C$_{\rm 1}$ & C$_{\rm s}$\\          
7 & D$_{\rm 5h}$ & C$_{\rm 3v}$ & C$_{\rm 2}$   &57 & C$_{\rm 1}$ & C$_{\rm 1}$ & C$_{\rm 1}$    & 107 & C$_{\rm s}$ & C$_{\rm 1}$ & C$_{\rm 1}$ \\          
8 & D$_{\rm 2d}$ & C$_{\rm s}$ & D$_{\rm 3d}$   &58 & C$_{\rm 1}$ & C$_{\rm s}$ & C$_{\rm 1}$    & 108 & C$_{\rm s}$ & C$_{\rm s}$ & C$_{\rm 1}$ \\            
9 & C$_{\rm 2v}$ & D$_{\rm 3h}$ & C$_{\rm s}$   &59 & C$_{\rm s}$ & C$_{\rm 1}$ & C$_{\rm 1}$    &  109 & C$_{\rm 3v}$ & C$_{\rm 1}$ & C$_{\rm 1}$\\           
10 & C$_{\rm 3v}$ & D$_{\rm 4d}$ & D$_{\rm 3h}$ &60 & C$_{\rm s}$ & C$_{\rm s}$ & C$_{\rm 1}$    &  110 & C$_{\rm 2v}$ & C$_{\rm 1}$ & C$_{\rm 1}$ \\          
11 & C$_{\rm 2v}$ & C$_{\rm 2}$ & C$_{\rm 2}$   &61 & C$_{\rm 3v}$ & C$_{\rm s}$ & C$_{\rm 2v}$  &  111 & C$_{\rm 1}$ & C$_{\rm 1}$ & C$_{\rm 1}$ \\           
12 & C$_{\rm 5v}$ &  C$_{\rm 2}$ & D$_{\rm 3h}$ &62 & C$_{\rm s}$ & C$_{\rm 1}$ & C$_{\rm 1}$    &  112 & C$_{\rm 2v}$ & C$_{\rm 1}$ & C$_{\rm 1}$ \\         
13 & I$_{\rm h}$ & C$_{\rm 1}$ & C$_{\rm s}$    &63 & C$_{\rm 2v}$ & C$_{\rm s}$ & C$_{\rm s}$   &  113 & T & C$_{\rm 2}$ & D$_{\rm 2}$ \\                   
14 & C$_{\rm 3v}$ & C$_{\rm 2v}$ & C$_{\rm s}$  &64 & C$_{\rm 2v}$ & C$_{\rm 1}$ & C$_{\rm s}$   &  114 & C$_{\rm 3}$ & C$_{\rm 2v}$ & C$_{\rm 2}$ \\         
15 & D$_{\rm 6d}$ & C$_{\rm 2v}$ & C$_{\rm s}$  &65 & C$_{\rm 1}$ & C$_{\rm 1}$ & C$_{\rm 1}$    &  115 &  C$_{\rm s}$ & C$_{\rm 1}$ & C$_{\rm s}$ \\         
16 & C$_{\rm 2v}$ & D$_{\rm 3h}$ & C$_{\rm 2v}$ &66 & C$_{\rm s}$ & C$_{\rm 2}$ & C$_{\rm 1}$    &  116 & C$_{\rm s}$ & O$_{\rm h}$ & C$_{\rm 3v}$ \\        
17 & T$_{\rm d}$ & D$_{\rm 4d}$ & C$_{\rm s}$   &67 & C$_{\rm s}$ & C$_{\rm 1}$ & C$_{\rm 1}$    &  117 & C$_{\rm s}$ & C$_{\rm s}$ & C$_{\rm 1}$\\            
18 & C$_{\rm 4v}$ & C$_{\rm 2v}$ & C$_{\rm s}$  &68 & C$_{\rm s}$ & C$_{\rm 1}$ & C$_{\rm 1}$    &  118 & C$_{\rm 2v}$ & C$_{\rm 1}$ & C$_{\rm 1}$\\          
19 & D$_{\rm 5h}$ & D$_{\rm 4d}$ & C$_{\rm 2v}$ &69 & C$_{\rm 2}$ & C$_{\rm 1}$ & C$_{\rm 1}$    &  119 & C$_{\rm 1}$ & C$_{\rm 1}$ & C$_{\rm 1}$\\           
20 & D$_{\rm 3d}$ & D$_{\rm 2}$ & D$_{\rm 2h}$  &70 & C$_{\rm 1}$ & C$_{\rm 1}$ & C$_{\rm 1}$    &  120 & C$_{\rm 1}$ & C$_{\rm 1}$ & C$_{\rm 1}$\\            
21 & C$_{\rm s}$ & C$_{\rm 1}$ & C$_{\rm s}$    &71 & C$_{\rm 2v}$ & C$_{\rm 2v}$ & C$_{\rm 1}$  &  121 & C$_{\rm 1}$ & C$_{\rm 1}$ & C$_{\rm 1}$\\            
22 & C$_{\rm 1}$ & C$_{\rm s}$ & D$_{\rm 6h}$   &72 & C$_{\rm s}$ & C$_{\rm s}$ & C$_{\rm s}$    &  122 & C$_{\rm 1}$ & C$_{\rm 1}$ & C$_{\rm 1}$\\           
23 & C$_{\rm 2v}$ & C$_{\rm s}$ & C$_{\rm s}$   &73 & C$_{\rm 2v}$ & C$_{\rm 1}$ & C$_{\rm s}$   &  123 &C$_{\rm s}$ &C$_{\rm 1}$ & C$_{\rm 1}$\\            
24 & C$_{\rm 2}$ & C$_{\rm s}$ & C$_{\rm 3v}$   &74 & C$_{\rm s}$ & C$_{\rm 5v}$ & C$_{\rm s}$   &  124 & C$_{\rm s}$ & C$_{\rm 1}$ & C$_{\rm s}$\\          
25 & C$_{\rm 2}$ & C$_{\rm 1}$ & C$_{\rm 2v}$   &75 & D$_{\rm 5h}$ & C$_{\rm 2v}$ & C$_{\rm s}$  &  125 & C$_{\rm 1}$ & C$_{\rm 1}$ & C$_{\rm 1}$\\         
26 & C$_{\rm 1}$ & C$_{\rm s}$ & C$_{\rm s}$    &76 & C$_{\rm s}$ & C$_{\rm s}$ & C$_{\rm s}$    &  126 & C$_{\rm s}$ & C$_{\rm 1}$ & C$_{\rm 1}$\\      
27 & C$_{\rm s}$ & C$_{\rm 2}$ & C$_{\rm s}$    & 77 & C$_{\rm 2v}$ & C$_{\rm s}$ & C$_{\rm 2v}$&127 & C$_{\rm 1}$ & C$_{\rm 1}$ & C$_{\rm 1}$\\   
28 & C$_{\rm s}$ & C$_{\rm 2}$ & C$_{\rm 2v}$   & 78 & C$_{\rm 2v}$ & C$_{\rm s}$ & C$_{\rm 1}$ &128 & C$_{\rm s}$ & C$_{\rm 1}$ & C$_{\rm s}$\\ 
29 & C$_{\rm 2}$ & C$_{\rm s}$ & C$_{\rm 2}$    & 79 & O$_{\rm h}$ & C$_{\rm s}$ & C$_{\rm 1}$  &129 & C$_{\rm 1}$ & C$_{\rm s}$ & C$_{\rm s}$\\   
30 &  C$_{\rm 3v}$ & C$_{\rm 1}$ & C$_{\rm 2}$  & 80 & C$_{\rm s}$ & O & C$_{\rm s}$            &130 & C$_{\rm s}$ & C$_{\rm s}$& C$_{\rm s}$\\    
31 & C$_{\rm 3}$ & C$_{\rm 1}$ & C$_{\rm 3v}$   & 81 & C$_{\rm 2v}$ & C$_{\rm s}$ & C$_{\rm 1}$ &131 & C$_{\rm 2v}$ & C$_{\rm 1}$ & C$_{\rm 1}$\\  
32 & D$_{\rm 2d}$ & C$_{\rm 3}$ & C$_{\rm 2v}$  & 82 & C$_{\rm s}$ & C$_{\rm 3}$ & C$_{\rm 2v}$ &132 & C$_{\rm s}$ & C$_{\rm 1}$ & C$_{\rm 1}$\\   
33 & C$_{\rm 2}$ & C$_{\rm 1}$ & C$_{\rm 1}$    & 83 & C$_{\rm s}$ & C$_{\rm s}$ & C$_{\rm s}$  &133 & C$_{\rm 2v}$ & C$_{\rm 1}$ & C$_{\rm 1}$\\  
34 & T$_{\rm d}$ & C$_{\rm 3}$ & C$_{\rm s}$    & 84 & C$_{\rm s}$ & C$_{\rm s}$ & C$_{\rm s}$  &134 & C$_{\rm s}$ & C$_{\rm s}$ & C$_{\rm s}$ \\  
35 & C$_{\rm 2v}$ & D$_{\rm 3}$ & C$_{\rm 2v}$  & 85 & C$_{\rm s}$ & C$_{\rm s}$ & C$_{\rm s}$  &135 & C$_{\rm s}$ & C$_{\rm 1}$ & C$_{\rm 1}$\\   
36 & C$_{\rm 2v}$ & C$_{\rm s}$ & D$_{\rm 2}$   & 86 & C$_{\rm s}$ & C$_{\rm s}$ & C$_{\rm s}$  &136 & C$_{\rm s}$ & C$_{\rm 1}$ & C$_{\rm 1}$\\   
37 & C$_{\rm 2v}$ & C$_{\rm s}$ & C$_{\rm 2}$   & 87 & C$_{\rm 2}$ & C$_{\rm 1}$ & C$_{\rm 1}$  &137 & C$_{\rm 2v}$ & C$_{\rm 1}$ & C$_{\rm 1}$\\  
38 & O$_{\rm h}$ & D$_{\rm 4h}$ & C$_{\rm s}$   & 88 & C$_{\rm 1}$ &  C$_{\rm 1}$ & C$_{\rm 1}$ &138 & C$_{\rm s}$ & C$_{\rm 1}$ & C$_{\rm 1}$\\   
39 &  D$_{\rm 3}$ & C$_{\rm s}$ & C$_{\rm 4v}$  & 89 & C$_{\rm 2}$ & C$_{\rm s}$ & C$_{\rm s}$  &139 & C$_{\rm 2v}$ & C$_{\rm 1}$ & C$_{\rm 1}$\\  
40 & D$_{\rm 3}$ & C$_{\rm 1}$ & C$_{\rm 2}$    & 90 & C$_{\rm 1}$ & C$_{\rm 1}$ & C$_{\rm 1}$  &140 & O$_{\rm h}$ & C$_{\rm s}$ & C$_{\rm s}$\\   
41 & C$_{\rm 1}$ & C$_{\rm 1}$ & C$_{\rm 1}$    & 91 & C$_{\rm 1}$ & C$_{\rm 1}$ & C$_{\rm 1}$  &141 & C$_{\rm 2v}$ & C$_{\rm 2v}$ & C$_{\rm s}$\\  
42 & D$_{\rm 4}$ & T$_{\rm d}$ & C$_{\rm 1}$    & 92 & C$_{\rm 1}$ & C$_{\rm 2}$ & C$_{\rm 1}$  &142 & C$_{\rm s}$ & C$_{\rm s}$ & C$_{\rm s}$ \\  
43 & D$_{\rm 2}$ & C$_{\rm 1}$ & C$_{\rm 1}$    & 93 & C$_{\rm 1}$ & C$_{\rm 1}$ & C$_{\rm 1}$  &143 & C$_{\rm s}$ & C$_{\rm 2v}$ & C$_{\rm s}$\\  
44 & C$_{\rm s}$ & C$_{\rm 2}$ & C$_{\rm 1}$    & 94 & C$_{\rm 1}$ & C$_{\rm 1}$ & C$_{\rm 1}$  &144 & C$_{\rm s}$ & C$_{\rm s}$ & C$_{\rm s}$\\   
45 & C$_{\rm s}$ & C$_{\rm 1}$ & C$_{\rm 1}$    & 95 & C$_{\rm 1}$ & C$_{\rm 1}$ & C$_{\rm 1}$  &145 & C$_{\rm 1}$ & C$_{\rm s}$ & C$_{\rm 1}$\\   
46 & C$_{\rm 3}$ & C$_{\rm s}$ & C$_{\rm 1}$    & 96 & C$_{\rm 1}$ & C$_{\rm 1}$ & C$_{\rm 1}$  &146 & D$_{\rm 5h}$ & C$_{\rm s}$ & C$_{\rm 1}$\\  
47 & C$_{\rm 1}$ & C$_{\rm 2}$ & C$_{\rm 1}$    & 97 & C$_{\rm 2}$ & C$_{\rm 1}$ & C$_{\rm 1}$  &147 & C$_{\rm 1}$ & C$_{\rm 1}$ & C$_{\rm 1}$\\   
48 & C$_{\rm 1}$ & D$_{\rm 2d}$ & C$_{\rm 1}$   & 98 & C$_{\rm 1}$ & C$_{\rm 1}$ & C$_{\rm 1}$  &148 & C$_{\rm 1}$ & C$_{\rm 1}$ & C$_{\rm 1}$\\   
49 & C$_{\rm 1}$ & C$_{\rm 1}$ & C$_{\rm s}$    & 99 & C$_{\rm 2}$ & C$_{\rm 1}$ & C$_{\rm 1}$  &149 & C$_{\rm 1}$ & C$_{\rm 1}$ & C$_{\rm 1}$\\   
50 & C$_{\rm 1}$ & C$_{\rm 1}$ & C$_{\rm 1}$    & 100 & C$_{\rm 1}$ & C$_{\rm 1}$ & C$_{\rm 1}$ &150 & C$_{\rm 1}$ & C$_{\rm 1}$ & C$_{\rm 1}$\\   
    
\hline\hline                                     
\end{tabular}                                  
\label{tab01}                                  
\end{center}                                   
\end{table}

\unitlength1cm                                 

\unitlength1cm                                 

\unitlength1cm                                 

\unitlength1cm                                 

\unitlength1cm

\unitlength1cm

\unitlength1cm

\unitlength1cm

\end{document}